\documentclass[review]{elsarticle}

\usepackage{lineno,hyperref}
\usepackage{amsfonts}
\usepackage{amsmath,amssymb}
\usepackage{multirow}
\usepackage{xcolor}
\modulolinenumbers[5]

\journal{ }

\begin{document}

\begin{frontmatter}
\linenumbers

\title{On the influence of Maxwell--Chern--Simons electrodynamics in nuclear fusion involving electronic and muonic molecules}


\author[mymainaddress]{Francisco Caruso}\corref{mycorrespondingauthor}
\cortext[mycorrespondingauthor]{Corresponding author}
\ead{ }

\author[mysecondaryaddress]{Vitor Oguri}
\author[mysecondaryaddress]{Felipe Silveira}
\author[mymainaddress]{Amos Troper}

\address[mymainaddress]{Centro Brasileiro de Pesquisas F\'{\i}sicas -- Rua Dr.~Xavier Sigaud, 150, 22290-180, Urca, Rio de Janeiro, RJ, Brazil}
\address[mysecondaryaddress]{Instituto de F\'{\i}sica Armando Dias Tavares, Universidade do Estado do Rio de Janeiro -- Rua S\~ao Francisco Xavier, 524, 20550-900, Maracan\~a, Rio de Janeiro, RJ, Brazil}

\begin{abstract}
New results recently obtained (\textit{Annals of Physics} (New York) a.n.~168943) established some non-relativistic ground state solutions for three-body molecules interacting through a Chern--Simons model. Within this model, it was argued that Chern--Simons potential should not help improve the fusion rates by replacing electrons with muons, in the case of particular muonic molecules. This achievement motivated us to investigate quantitatively whether or not the Maxwell--Chern--Simons electrodynamics could influence positively, for example, the probability of having a muon-catalyzed fusion; its contribution to electronic molecules is also considered in this letter. The principal factors related to the probability of elementary nuclear fusion are therefore numerically calculated and compared with their analogs admitting other forms of interaction like $-1/\rho$ and $\ln (\rho)$. The analysis carried on here confirms that one should not expect a significant improvement in nuclear fusion rates in the case of muonic molecules, although, surprisingly, the same is not true for electronic molecules, compared with other theoretical predictions. Numerical predictions for the fusion rates for $ppe$, $pp\mu$, $dde$ and $dd\mu$ molecules are given as well as the predicted value for the tunneling rate for these molecules.

\end{abstract}


\end{frontmatter}

\newpage


Recent studies of elementary processes have shown that muon-catalyzed fusion can experience a significant gain when observed within a purely two-dimensional space~\cite{caruso2019planar,felipeepjd}. However, the actual nature of interaction potential between electric charge carriers inside the molecules is still an open issue ~\cite{felipe2}. A possible answer could be found within the framework of the so called Maxwell--Chern--Simons theory, where the electromagnetic interaction allows the photon to acquire a non-trivial topological mass ($m_\gamma$) without any conflict with its inherent gauge symmetry. This trend is due in part to the understanding that there is a formal connection between planar quantum electrodynamics and the Chern--Simons theory, as shown in~\cite{felipeannals} and references therein.

It is a general belief that nuclear fusion can be favored with the use of muonic molecules. The main argument for this is that since the muon mass is about 200 times greater than that of the electron, The muonic Bohr radius will be also roughly 200 times smaller, possibly favoring the fusion process. Previous analyzes~\cite{felipeannals} have already shown that this fact is not true for molecules governed by the Chern--Simons interaction. In this letter, the same kind of analysis made in reference~\cite{felipeepjd} is carried on in the light of the new results of reference~\cite{felipeannals}. The purpose is to find out if the electrodynamics of Maxwell--Chern--Simons can exert a positive influence on ordinary molecules and on muon-catalyzed fusion or not.

It is well known that the Chern--Simons potential depends on the effective topological mass of the photon $m_{\gamma}$, which will be taken as a fraction of the electronic mass, $m_\gamma = \lambda m_e$. All the predictions of this letter were get for two different values of $m_{\gamma}$, corresponding to the choices $\lambda=0.2\times10^{-3}$ and $\lambda=0.2\times10^{-5}$.

Fusion rates can be estimated as the probability of finding the two molecular nuclei very close together, as in Ref.~\cite{felipeepjd}. To calculate this probability for an internuclear separation inside a region limited by points $a$ and $b$, in 3$D$, is given by

\begin{equation}
  \int_{a}^{b} |\psi(\rho)|^2\rho^2\mbox{d}\rho
\end{equation}
and, in 2D, is given by
\begin{equation}
  \int_{a}^{b} |\psi(\rho)|^2\rho\,\mbox{d}\rho
\end{equation}

To facilitate the comparison between our predictions and other previous calculations, this separation was fixed at 10~fm, and the results are shown in Table~\ref{prob}.
\renewcommand{\arraystretch}{0.9}
\begin{table}[hb]
 \caption{Theoretical predictions for the probabilities of finding the molecules with a small inter nuclear separation. The first Chern--Simons (CS) column correspond to the choice of $\lambda=0.2\times10^{-3}$ and the second one to $\lambda=0.2\times10^{-5}$.} \label{prob}
 \vspace*{0.2cm}
  \centering
  \begin{tabular}{c c c c c}
\hline
\hline
      Molecule      &\hspace{0.3cm} CS   &\hspace{0.3cm}  CS  &\hspace{0.3cm} $ln(\rho)$  &\hspace{0.3cm}  $-1/\rho$         \\ \hline
      $ppe$         &\hspace{0.3cm} $3.47\times10^{-6}$ &\hspace{0.3cm} $2.68\times10^{-5}$     &\hspace{0.3cm} $7.23\times10^{-6}$   &\hspace{0.3cm}  $2.4\times10^{-84}$         \\
      $dde$         &\hspace{0.3cm} $2.27\times10^{-6}$  &\hspace{0.3cm} $1.6\times10^{-5}$     &\hspace{0.3cm} $4.56\times10^{-6}$   &\hspace{0.3cm}  $1.7\times10^{-88}$         \\
      $pp\mu$       &\hspace{0.3cm} $2.98\times10^{-6}$ &\hspace{0.3cm} $2.42\times10^{-5}$     &\hspace{0.3cm} $9.3\times10^{-3}$   &\hspace{0.3cm}  $2.9\times10^{-9}$         \\
      $dd\mu$       &\hspace{0.3cm} $9.54\times10^{-5}$ &\hspace{0.3cm} $8.97\times10^{-4}$    &\hspace{0.3cm} $1.6\times10^{-2}$   &\hspace{0.3cm}  $1.2\times10^{-10}$        \\ \hline
      \hline
\end{tabular}
\end{table}
\renewcommand{\arraystretch}{1}


The second parameter that should be investigated is how the tunneling coefficient ($T$), which changes according to the choice of the interaction potential.

From the Schr\"{o}dinger equation that describes each of these molecules, we have that the transmission coefficient~\cite{razavy} for a particle tunneling through a potential barrier, in dimensionless unity, is given by the expression
\begin{equation}\label{tunnell}
  \exp\left(-2\int_{\rho_1}^{\rho_2}\mbox{d}\rho \sqrt{V(\rho)-\varepsilon}\right)
\end{equation}
\noindent where $V(\rho)$ is the respective effective potential, and $\rho_1$ and $\rho_2$ are the two classical turning points for the potential barrier. Eq.~(\ref{tunnell}) gives us a semi-classical estimation for $T$ (the tunneling coefficients), as shown in the Table~\ref{tabelatun}.
\renewcommand{\arraystretch}{1.0}
\begin{table}[ht]
  \caption{Tunnelling coefficients ($T$) calculated from Eq.~\ref{tunnell} for different molecules with three different interaction potentials. The first Chern--Simons (CS) column correspond to the choice of $\lambda=0.2\times10^{-3}$ and the second one to $\lambda=0.2\times10^{-5}$.}\label{tabelatun}
  \vspace*{0.2cm}
  \begin{center}
  \begin{tabular}{c c c c c}
    \hline
    \hline
      Molecule    &\hspace{0.3cm}  $T^{(CS)}$&\hspace{0.3cm}  $T^{(CS)}$&\hspace{0.3cm}  $T^{ln(\rho)}$&\hspace{0.3cm}  $T^{-1/\rho}$     \\ \hline
      $ppe$       &\hspace{0.3cm}  $8.54\times10^{-3}$   &\hspace{0.3cm}  $9.87\times10^{-2}$   &\hspace{0.3cm}  $9.59\times10^{-8}$         &\hspace{0.3cm}  $6.89\times10^{-71}$           \\
      $dde$       &\hspace{0.3cm}  $7.99\times10^{-3}$   &\hspace{0.3cm}  $9.92\times10^{-2}$   &\hspace{0.3cm}  $8.47\times10^{-8}$         &\hspace{0.3cm}  $1.44\times10^{-97}$           \\
      $pp\mu$     &\hspace{0.3cm}  $9.46\times10^{-3}$   &\hspace{0.3cm}  $9.89\times10^{-2}$   &\hspace{0.3cm}  $1.56\times10^{-2}$         &\hspace{0.3cm}  $1.92\times10^{-5}$         \\
      $dd\mu$     &\hspace{0.3cm}  $9.39\times10^{-3}$   &\hspace{0.3cm}  $9.88\times10^{-2}$   &\hspace{0.3cm}  $8.03\times10^{-4}$         &\hspace{0.3cm}  $1.73\times10^{-7}$         \\ \hline

      \hline
  \end{tabular}
  \end{center}
\end{table}
\renewcommand{\arraystretch}{1}

Let us present now some discussions concerning the above results.

First, it is important to point out that the analyzes carried out in this work are estimates, as they do not take into account the completeness of the processes involved in this type of fusion. A direct inspection of the results in Table~\ref{prob} shows that the Chern--Simons potential could exert a significant influence, presenting an even better rate than the logarithmic potential in the case of electronic molecules, although we don't see that same improvement when we look at muonic molecules as usually expected. This is due to the influence of the topological mass of the photon. Indeed, changing from electron to muon does not have the same influence on the size of the molecule as the $\ln{\rho}$ or the $-1/\rho$ case (a consequence of its small value, \textit{i.e.}, $0.2 \times 10^{-3} \leq \lambda \leq 0.2 \times 10^{-5}$), as already shown in Ref.~\cite{felipeannals}.

Comparing the results of Table~\ref{prob} with those given in Ref.~\cite{caruso2019planar}, it comes out a fusion rate for $pp\mu$ molecule which is of the order of $10^{5}$ times greater than the predicted for 3D. Meanwhile, our prediction for the fusion rate in the case of $ppe$ is of the same order of magnitude for the Chern--Simons and logarithmic potentials, considering $\lambda \sim 0.2 \times 10^{-5}$.

Concerning tunneling rates, for the $pp\mu$ molecule, it was shown in Refs.~\cite{caruso2019planar,felipeepjd} that it is amplified by a factor $\simeq 10^{4}$ for the $\ln (\rho)$ potential in 2D compared with the 3D result. In this letter it is shown that, for CS interaction, this rate is of the same order of magnitude.
\newpage

The surprising result comes from $ppe$ molecule with Cs interaction. In this case, the gain in tunneling rate is still bigger, being of the order of $10^{6}$ (Table~\ref{tunnell}).

Initially, the evaluation of these theoretical results show that muon-catalyzed fusion is not facilitated when we are in a domain of the Maxwell--Chern--Simons electrodynamics, when compared to the logarithmic case. However, we observed a new possibility, hitherto unheard of, which is the ability to perform low temperature fusions without the need to replace the electron with the muon, as we can see in Table~\ref{prob}, the electronic molecules presents an even greater probability of fusion than its analogs with the logarithmic potential.


In conclusion, the adiabatic model for nuclear fusion developed in Refs.~\cite{caruso2019planar,felipeepjd} suggests that fusion rates will be significatively enhanced in 2D, no matter if the inter-molecular potential is modeled by $\ln(\rho)$ or by the Maxwell--Chern--Simons one. This is not a trivial result since the two potentials have similar behavior just for quite small values of $\rho$. Actually, we have shown in this letter that both potentials give rise to quantitative different predictions in the case of $ppe$. The contribution of Maxwell--Chern--Simons potential to fusion rates of $pp\mu$ is one or two orders of magnitudes less than the equivalent rates predicted adopting the logarithmic potential.

\section*{Acknowledgment}

One of us (FS) was financed in part by the Coordena\c{c}\~{a}o de Aperfei\c{c}oamento de Pessoal de N\'{\i}vel Superior -- Brazil (CAPES), Finance Code 001.

\end{document}